\documentstyle[12pt]{article}

\makeatletter
\renewcommand\thesubsection{\thesection.\@arabic\c@subsection}
\makeatother

\newcommand {\beq}{\begin{equation}}
\newcommand {\eeq}{\end{equation}}
\newcommand {\beqa}{\begin{eqnarray}}
\newcommand {\eeqa}{\end{eqnarray}}         %Equation version
\newcommand {\beqs}{\begin{eqnarray*}}
\newcommand {\eeqs}{\end{eqnarray*}}
\newcommand {\bds}{\begin{displaymath}}
\newcommand {\eds}{\end{displaymath}}
\newcommand {\n}{\nonumber\\}

\newcommand {\bebb}{\begin{thebibliography}}
\newcommand {\eebb}{\end{thebibliography}}      %Reference version

\begin{document}

\vskip 1cm

\begin{center}
%\title
{\large\bf Analytic solutions of the Teukolsky equation for massless perturbations of any spin in de Sitter background}

\vspace{0.5cm}

%\author{
{\large Yao-Zhong Zhang}
\vskip.1in

{\em School of Mathematics and Physics, The University of Queensland, \\
Brisbane, Qld 4072, Australia}

\end{center}

\date{}

%\maketitle

%\vspace{2cm}

\begin{abstract}
We present analytic solutions to the Teukolsky equation for massless perturbations of any spin in the 
4-dimensional de Sitter background. The angular part of the equation fixes the separation constant
to a discrete set and its solution is given by hypergeometric polynomials. For the radial part, we derive 
analytic power series solution which is regular at the poles and determine a transcendental function 
whose zeros give the characteristic values of the wave frequency.  We study the existence of explicit
polynomial solutions to the radial equation and obtain two classes of singular closed-form solutions, 
one with discrete wave frequencies and the other with continuous frequency spectra.  
\end{abstract}

\vskip.1in

{\tt PACS numbers}: 03.65.Ge, 04.30.Nk, 04.20.Cv.

%\noindent \emph{Keywords}: Teukolsky master equation, perturbation of black hole, three-term recurrence relation, polynomial solution.

%\maketitle

%\end{frontmatter}

\section{Introduction}
Perturbations to known solutions of Einstein's equations by various types of fields, such as scalar,
neutrino, electromagnetic, and gravitational fields, have been extensively studied in the literature. 
One of the important solutions is the de Sitter spacetime. It is the simplest model of spacetime with 
a non-zero cosmological constant and is relevant to inflation and the physics of very early universe 
\cite{Guth81,Albrecht82}. Experimental evidences and astronomical observations, combining with the theory of
inflation, suggest that our universe is expanding in an accelerated rate and may approach
the de Sitter geometry asymptotically \cite{Perlmutter99}. It has also recently been conjectured that 
there exists a holographic duality between quantum gravity in (anti-)de Sitter spacetime and certain
conformal field theory on the boundary of that space (the so-called AdS/CFT correspondence) 
\cite{Strominger01,Mazur01}. Therefore, it is certainly of interest to investigate perturbations in 
the static region of the de Sitter spacetime between the origin and the cosmological horizon. 

In this paper we discuss gauge- and tetrad-invariant first order massless perturbations of any spin
in the de Sitter spacetime. The dynamics of these perturbations in four dimensions
can be conveniently described by the Teukolsky master differential equation \cite{Teukolsky72}. 
The radial and angular parts of the Teukolsky perturbation equation
are separable thanks to the spherical symmetry of the de Sitter background spacetime.  
We provide representations of analytic solutions to both the angular and radial equations. We show that 
the angular part of the equation fixes the separation 
constant to a discrete set and its solution is given by hypergeometric polynomials. We obtain 
power series solution of the radial equation which is regular at the poles and determine a transcendental
function whose zeros give the characteristic values of the wave frequency. 
Furthermore, explicit polynomial solutions
of the radial equation are studied by applying the general procedure and results of the present 
author in \cite{Zhang12}. Two classes of singular closed-form solutions to the radial equation are found:
one yields discrete complex frequencies while the frequency of the other class has a continuous spectrum.
Let us mention that polynomial solutions to the Teukolsky master equation with a continuous frequency 
spectrum also were studied previously for perturbations in the Kerr background \cite{Borissov09}.

The outline of the present work is as follows. In section \ref{master} we briefly review the Teukolsky master
perturbation equation and the separation of its variables. Section \ref{angular} deals with the angular
eigenvalue problem. The regularity of the angular function at the poles is used to fix the separation
constant. In section \ref{radial} we study the radial eigenvalue problem. We follow a procedure which is 
based on the application of the mathematical theorems on solutions of three-term recurrence relation 
\cite{Gautschi67,Leaver86}. (A similar procedure has recently used to find entire function
solutions of the Rabi model \cite{Moroz12} and its generalizations \cite{Zhang14,Zhang20}.)
In section \ref{polynomial} we examine the existence of explicit, polynomial solutions to the radial 
equation. We present two classes of a total five families of closed-form solutions, 
all of them are singular at the poles.  We conclude the work with a summary in section \ref{summary}.

\section{Teukolsky master equation for the perturbations}\label{master}
We consider gauge- and tetrad-invariant first order massless perturbations of any spin in the 
de Sitter background. In static coordinates, the de Sitter metric takes the form
\beq
ds^2=N^2 dt^2-N^{-2} dr^2-r^2\left(d\theta^2+\sin^2\theta d\phi^2\right)
\eeq
where $N=\sqrt{1-H^2 r^2}$ denotes the lapse function. Note that there is a cosmological horizon at 
$r=r_H=\frac{1}{H}$ and the spacetime region which can be accessed is the ball of radius $r_H$, 
centered at the origin. The Teukolsky master equation governing the perturbations is given by \cite{Bini11}
\beqa
&&\left\{\frac{1}{2N^2}\frac{\partial^2}{\partial t^2}-\frac{N^2}{2}\frac{\partial^2}{\partial r^2}
   -\frac{1}{2r^2}\frac{\partial^2}{\partial \theta^2}-\frac{1}{2r^2\sin^2\theta}\frac{\partial^2}{\partial\phi^2}
   +\frac{s}{rN^2}\frac{\partial}{\partial t}+\frac{(s+1)(1-2N^2)}{r}\frac{\partial}{\partial r}   \right.\n
&&~~~~\left.-\frac{\cot\theta}{2r^2}\frac{\partial}{\partial \theta}-\frac{is\cos\theta}{r^2\sin^2\theta}\frac{\partial}{\partial \phi}
   +\frac{1}{2r^2}\left[(s+1)\left(3s+2-2N^2(2s+1)\right)+\frac{s^2}{\sin^2\theta}\right]\right\}\psi=0.\n
\eeqa
The parameter $s$ is called the spin weight of the field, and is given by $s = \pm 2$ for 
gravitational perturbations, $s = \pm 1$ for electromagnetic perturbations, $s = \pm \frac{1}{2}$
for massless neutrino perturbations, and $s = 0$ for scalar perturbations.
The variables in the Teukolsky master equation can be separated by using the ansatz 
\beq
\psi(t,r,\theta,\phi)=e^{-i\omega t} e^{im\phi} R(r)\,\Theta(\theta),
\eeq
where $m$ is the azimuthal parameter. 
Then one obtains the following angular and radial equations,
\beqa
&&\frac{1}{\sin\theta}\frac{d}{d\theta}\left(\sin\theta\frac{d\Theta(\theta)}{d\theta}\right)
  +\left[\Lambda-s^2-\frac{(m+s\cos\theta)^2}{\sin^2\theta}\right]\Theta(\theta)=0,\n 
&&Q^{-s}\frac{d}{dr}\left(Q^{s+1}\frac{dR(r)}{dr}\right)\n
&&~~~~+\left[\frac{\omega^2r^2+2is\omega r}{N^2}
  +2N^2(s+1)(2s+1)-\Lambda-(s+1)(3s+2)\right]R(r)=0,
\eeqa
where $Q=r^2N^2$. The complex parameters $\omega$ and $\Lambda$ are the wave frequency and
separation constant, respectively.
The sign of the imaginary part ${\rm Im}(\omega)$ of the complex wave frequency $\omega$
determines whether the solution is stable (decaying
in time with ${\rm Im}(\omega)<0$) or unstable (growing in time with ${\rm Im}(\omega)>0$). 
Stationary modes are characterized by ${\rm Im}(\omega)=0$. Obviously the angular equation is
independent of the frequency $\omega$.

\section{The angular eigenvalue problem}\label{angular}
The angular functions $\Theta(\theta)$ are required to be regular at the poles $\theta = 0$ and 
$\theta=\pi$. These boundary conditions pick out a discrete set of 
$\Lambda$, as shown below. In terms of the variable $x=\cos\theta$, the angular equation becomes
\beq
(1-x^2)\,\Theta''-2x\,\Theta'+\left[\Lambda-\frac{(s+m)^2/2}{1-x}-\frac{(s-m)^2/2}{1+x}\right]
   \Theta=0,~~~~~~|x|\leq 1. 
\eeq
The solutions to this equation at the regular singularities $x=\pm 1$ can be found in the usual way. If
\beq
\lim_{x\rightarrow 1}\Theta \sim (1-x)^{k_1},~~~~~~\lim_{x\rightarrow -1}\Theta \sim (1+x)^{k_2},
\eeq
then
%\beq
$k_1=\pm\frac{1}{2}(s+m)$%,~~~~~~~~
and $k_2=\pm \frac{1}{2}(s-m)$.
%\eeq
The physically meaningful solutions to the angular equation are those that are regular at $x=\pm 1$
(which correspond to $\theta=0$ and $\theta=\pi$, respectively), 
so the usual choices for $k_1$ and $k_2$ are $k_1=|s+m|/2$ and $k_2=|s-m|/2$. Setting
\beq
\Theta(x)=(1-x)^{{|s+m|}/{2}}(1+x)^{{|s-m|}/{2}}\,S(x)
\eeq
yields 
\beq
(x-1)(x+1)S''+[(a+b+1)x+|s+m|+|s-m|]S'+ab\,S=0,
\eeq
where
\beqa
a&=&\frac{1}{2}\left[|s+m|+|s-m|+1+\sqrt{4\Lambda+1}\right],\n
b&=&\frac{1}{2}\left[|s+m|+|s-m|+1-\sqrt{4\Lambda+1}\right].
\eeqa
Making a change of variable $x=2z-1$  converts the above into the 
hypergeometric differential equation for $S(z)$,
\beq
z(z-1)S''+\left[(a+b+1)z-(|s-m|+1)\right]S'+ab\,S=0,~~~~~0\leq z\leq 1.
\eeq
The local solution of this equation around $z=0$ is known as the hypergeometric function
\beq
S(z)=\sum_{n=0}^\infty Q_n\,z^n,
\eeq
where the coefficients $Q_n$ are functions of the separation constant $\Lambda$ and are
determined by the recurrence relation
\beq
Q_{n+1}=\frac{n(n+a+b)+ab}{(n+1)(n+1+|s-m|)}Q_n,~~~~~Q_0=1.
\eeq
If for some positive integer $N=0, 1, \cdots,$
\beq
N(N+a+b)+ab=0,\label{condition}
\eeq
then we have $Q_{N+1}=0$ and the hypergeometric function terminates 
to the hypergeometric polynomial of degree $N$. Under the condition (\ref{condition}) 
we obtain the exact solution of the angular equation,
\beq
\Theta(\theta)=(1-\cos\theta)^{|s+m|/2}(1+\cos\theta)^{|s-m|/2}\,
   S_N(\cos\theta),
\eeq
where $S_N$ is the polynomial in $\cos\theta$ of degree $N$: 
$S(\cos\theta)=\sum_{n=0}^N  Q_n(\Lambda)\left(\frac{1+\cos\theta}{2}\right)^n$.
The corresponding separation constant $\Lambda$ takes discrete values given by 
\beq
\Lambda=\frac{1}{4}(2N+|s-m|+|s+m|)(2N+|s-m|+|s+m|+2),~~~~~N=0,1,2,\cdots.\label{lambda}
\eeq
So the separation constant $\Lambda$ is fixed by the angular part of the Teukolsky equation alone!

\section{Three-term recurrence relation and the radial eigenvalue problem}\label{radial}
In this section we present analytic solution to the radial equation on $0\leq r\leq r_H$.
By the substitution
\beq
R(r)=Q^{-\frac{s+1}{2}}Y(r),
\eeq 
one can convert the radial equation into the normal Schr\"odinger form \cite{Bini11}
\beqa
&&\left[-\frac{d^2}{dr^2}+V(r)\right]Y(r)=0,\n
&&V(r)=\frac{c_0}{r^2}+\frac{c^-_H}{(r-r_H)^2}+\frac{c^+_H}{(r+r_H)^2}+
  \frac{d_0}{r} +\frac{d^-_H}{r-r_H}+\frac{d^+_H}{r+r_H},\label{schroedinger}
\eeqa
where $c_0=\Lambda,~~d_0=-2is\omega$ and 
\beqa
%c_0&=&\Lambda,~~~~~~~~d_0=-2is\omega,\n
c^\pm_H&=&\frac{1}{4}\left[(s\pm i\omega r_H)^2-1\right],\n
d^\pm_H&=&\mp\frac{1}{4r_H}\left(\omega^2r^2_H\mp 4is\omega r_H-s^2-2\Lambda+1\right).
\eeqa
Note in passing that $d_0+d^+_H+d^-_H=0$ and the separation constant $\Lambda$, fixed 
by the angular part, is given by (\ref{lambda}). 
Solutions of the above equation (\ref{schroedinger}) at the regular singularities
$r=0, \pm r_H$ can be found as follows. If
\beq
\lim_{r\rightarrow 0}Y\sim r^{\alpha_0},~~~~~\lim_{r\rightarrow r_H}Y\sim(r_H-r)^{\alpha^-_H},~~~~~
\lim_{r\rightarrow -r_H}Y\sim(r_H+r)^{\alpha^+_H}
\eeq
or equivalently
\beq
\lim_{r\rightarrow 0}R\sim r^{-s-1+\alpha_0},~~~~~
\lim_{r\rightarrow r_H}R\sim(r_H-r)^{-\frac{s+1}{2}+\alpha^-_H},~~~~~
\lim_{r\rightarrow -r_H}R\sim(r_H+r)^{-\frac{s+1}{2}+\alpha^+_H},
\eeq
then 
\beq
\alpha_0=\frac{1}{2}\left[1\pm \sqrt{4\Lambda+1}\right],~~~~~~
\alpha^-_H=\frac{1}{2}[1\pm(s- i\omega r_H)],~~~~~~
\alpha^+_H=\frac{1}{2}[1\pm(s+ i\omega r_H)]. \label{alpha0-+}
\eeq
%The signs have to be chosen such that the 
Note that solutions of the radial equation which are nonsingular at $r=0, \pm r_H$ correspond to the choice
\beq
\alpha_0=\frac{1}{2}\left[1+ \sqrt{4\Lambda+1}\right],~~~~~~
%\alpha^-_H=\frac{1}{2}[1\pm(s- i\omega r_H)],~~~~
\alpha^\mp_H=\frac{1}{2}[1+s\mp i\omega r_H],~~~~~~{\rm Im}(\omega)\geq 0. \label{alpha0-+nonsingular}
\eeq
Applying the substitution,
\beq
Y(r)=r^{\alpha_0}\left(r_H-r\right)^{\alpha^-_H}\left(r+r_H\right)^{\alpha^+_H}X(r),
\eeq
%and choosing 
%\beq
%\alpha_0(\alpha_0-1)=c_0,~~~~~\alpha^-_H(\alpha^-_H-1)=c^-_H,~~~~~\alpha^+_H(\alpha^+_H-1)=c^+_H,
%    \label{eqns-for-a0a-a+}
%\eeq
%with $\alpha_0, \alpha^\pm_H$ given by (\ref{alpha0+-}), 
we can transform (\ref{schroedinger}) into 
%\beq
% X''+2\left(\frac{\alpha_0}{r}+\frac{\alpha^-_H}{r-r_H}+\frac{\alpha^+_H}{r+r_H}\right)
%   X'+\frac{[(d^+_H-d^-_H)r_H+2(\alpha^+_H\alpha^-_H+\alpha_0\alpha^+_H+\alpha_0 \alpha^-_H)] r
%   -q}{r(r-r_H)(r+r_H)}\,X=0,
%\eeq
%\beq
% X''+\left(\frac{2\alpha_0}{r}+\frac{2\alpha^-_H}{r-r_H}+\frac{2\alpha^+_H}{r+r_H}\right)
%   X'+\frac{\alpha\beta\, r  -q}{r(r-r_H)(r+r_H)}\,X=0,\label{Heun-equation}
%\eeq
\beq
r(r^2-r^2_H)X''+\left[(\alpha+\beta+1)r^2+2\left(\alpha^-_H-\alpha^+_H\right)r_H\,r-2\alpha_0 r^2_H
   \right]X'+(\alpha\beta\,r-q)X=0,\label{Heun-equation}
\eeq
where
\beqa 
\alpha&=&\alpha_0+\alpha^+_H+\alpha^-_H,\n
\beta&=&\alpha_0+\alpha^+_H+\alpha^-_H-1,\n
q&=&2\alpha_0(\alpha^+_H-\alpha^-_H)r_H-d_0r^2_H.\label{alpha-beta-q}
\eeqa
%This is the Heun general equation with four regular singular points $0, \pm r_H, \infty$. 
%It is not hard to check that 
%$(d^+_H-d^-_H)r_H+2(\alpha^+_H\alpha^-_H+\alpha_0\alpha^+_H+\alpha_0 \alpha^-_H)$ 
%factorizes into $\alpha\beta$ with

We now make a change of variable, $\tilde{r}=r_H-r$. In terms of $\tilde{r}$ the physically accessible
spacetime region is the ball $0\leq\tilde{r}\leq r_H$, and (\ref{Heun-equation}) becomes
\beq
 X''+\left(\frac{2\alpha^-_H}{\tilde{r}}+\frac{2\alpha_0}{\tilde{r}-r_H}
   +\frac{2\alpha^+_H}{\tilde{r}-2r_H}\right)
   X'+\frac{\alpha\beta\, \tilde{r}  -\tilde{q}}{\tilde{r}(\tilde{r}-r_H)(\tilde{r}-2r_H)}\,X=0,
    \label{Heun-equation1}
\eeq
where $\tilde{q}=\alpha\beta\,r_H-q$.  This is the Heun general
equation with four regular singular points  $\tilde{r}=0, r_H, 2r_H, \infty$.
% which is of the same form as (\ref{Heun-equation}). 
We seek power series solution of (\ref{Heun-equation1}), 
\beq
X(\tilde{r})=\sum_{n=0}^\infty {K}_n(\omega)\,\tilde{r}^n\label{power-series-soln1}
\eeq
which is convergent in the ball $0\leq \tilde{r}\leq r_H$ (i.e. $0\leq r\leq r_H$, where the coefficients ${K}_n$ are functions of the wave frequency $\omega$. 
Substituting (\ref{power-series-soln1}) into (\ref{Heun-equation1}), we find 
that ${K}_n$ obey the 3-term recurrence relation
\beqa
&&{K}_1+{A}_0\, {K}_0=0,\n
&&{K}_{n+1}+{A}_n\,{K}_n+{B}_n\,{K}_{n-1}=0,~~~~~n\geq 1 
     \label{three-term-relation1}
\eeqa
with 
\beqa
{A}_n&=&-\frac{\alpha\beta\,r_H-q+n\left(3n-3+4\alpha_0+2\alpha^+_H+6\alpha^-_H\right)\,r_H}
   {2(n+1)(n+2\alpha^-_H)\,r^2_H}\n
%&=&-\frac{n(3n+2s+1)+L(4n+L+2s+1)}{2(n+1)(n+s+1-i\omega r_H)r_H},\n
{B}_n&=&\frac{(n-1+\alpha)(n-1+\beta)}{2(n+1)(n+2\alpha^-_H)\,r^2_H}.\label{A&B}
%    =\frac{(n+L)(n+L-1)}{2(n+1)(n+s+1-i\omega r_H)r^2_H}.
\eeqa
%Here we have introduced the notation $L=\frac{1}{2}\left[1+\sqrt{4\Lambda+1}-2i\omega r_H\right]$.
Solutions to the 3-term recurrence relation can be classified by applying the mathematical theorems 
in \cite{Gautschi67}. The characteristic equation of (\ref{three-term-relation1}) is given by
%\beq
$u^2-\frac{3}{2r_H}u+\frac{1}{2r^2_H}=0$.
%\eeq
This equation has two distinct roots $u_1=\frac{1}{2r_H}$ and $u_2=\frac{1}{r_H}$ and $|u_1|<|u_2|$. 
By the Perron theorem (i.e. Theorem 2.2 of \cite{Gautschi67}), 
there exist two linearly independent solutions ${K}_{n,1}$ and ${K}_{n,2}$ of
(\ref{three-term-relation1}) such that
\beq
\lim_{n\rightarrow\infty}\frac{{K}_{n+1,s}}{{K}_{n,s}}=u_s,~~~~~~s=1,2.\label{asymptotics}
\eeq
Thus  ${K}_n^{\rm min}\equiv {K}_{n,1}$ is a minimal solution of (\ref{three-term-relation1}), while 
the other solution $K_{n,2}$ is dominant. 

By the Pincherle theorem (i.e. Theorem 1.1 of \cite{Gautschi67}), the ratio of successive elements of
the minimal solution sequence ${ K}_n^{min}$ is expressible as continued fractions,
%Proceeding in the direction of increasing $n$, we have
\beq
R_{n}=\frac{{ K}_{n+1}^{min}}{{ K}_n^{min}}=-\frac{{B}_{n+1}}{~{A}_{n+1}-}\,
  \frac{{B}_{n+2}}{~{A}_{n+2}-}\,\frac{{B}_{n+3}}{~{A}_{n+3}-}\,\cdots,   
   \label{continued-fraction}
\eeq
which for $n=0$ gives
\beq
R_{0}=\frac{{ K}_{1}^{min}}{{ K}_0^{min}}=-\frac{{B}_{1}}{~{A}_{1}-}\,
   \frac{{B}_{2}}{~{A}_{2}-}\,\frac{{B}_{3}}{~{A}_{3}-}\,\cdots.   
   \label{continued-fraction1}
\eeq
Note that the ratio $R_0=\frac{{ K}_1^{min}}{{ K}_0^{min}}$ involves ${ K}_0^{min}$, 
although the above continued fraction expression is obtained from the 
2nd equation of (\ref{three-term-relation1}), i.e the recurrence (\ref{three-term-relation1}) for $n\geq 1$. 
However, for single-ended sequences such as those appearing in the infinite series expansion
(\ref{power-series-soln1}), the ratio $R_0=\frac{{ K}_1^{min}}{{ K}_0^{min}}$ of 
the first two terms of a minimal solution is unambiguously fixed by the first equation of 
the recurrence (\ref{three-term-relation1}), namely,
%\begin{widetext}
\beq
 R_0=-{A}_0=\frac{\alpha\beta\,r_H-q}{4\alpha^-_H\,r^2_H}. % =\frac{L(L+2s+1)}{2(s+1-i\omega r_H)r_H}.
   \label{continued-fraction2}
\eeq
%\end{widetext}
In general, the $R_0$ computed from the continued fraction (\ref{continued-fraction1}) 
can not be the same as that from (\ref{continued-fraction2}) for arbitrary values of
recurrence coefficients ${A}_n$ and ${B}_n$. As a result, general solutions to the recurrence 
(\ref{three-term-relation1}) are dominant and are usually
generated by simple forward recursion from a given value of ${ K}_0$. The resulting power series
(\ref{power-series-soln1}) will converge for $0\leq \tilde{r}< r_H$ but will diverge when $\tilde{r}=r_H$.
This is seen as follows. By d'Alemmbert's Ratio Test, the radius $\rho$ of convergence of the power series
expansion (\ref{power-series-soln1}) is given by
%\beq
$\rho^{-1}=\lim_{n\rightarrow\infty}\frac{{K}_{n+1}}{{K}_{n}}$.
%\eeq
It follows from (\ref{asymptotics}) that $\rho$ equals to $2r_H$ for the minimal solution sequence
$K_{n,1}$ and to $r_H$ for the dominant one $K_{n,2}$.
Thus the power series expansion generated by the dominant solution sequences is only convergent
inside the ball $0\leq {\tilde r}<r_H$ but not on the boundary $\tilde{r}=r_H$. 
The physically meaningful solutions to the radial equation are
those that are convergent at both $\tilde{r}=0$ and $\tilde{r}=r_H$ 
(which correspond to $r=r_H$ and $r=0$, respectively). 
This will happen only for certain characteristic values of the frequency $\omega$ so that equations 
(\ref{continued-fraction1}) and (\ref{continued-fraction2})
are both satisfied. Then the resulting solution sequence ${ K}_n$ will be purely minimal and 
the corresponding power series expansion (\ref{power-series-soln1}) 
 will be convergent for $0\leq \tilde{r}< 2r_H$, thus it converges at both singular points
$\tilde{r}=0$ and $\tilde{r}=r_H$.

Therefore, if we define the transcendental function $F(\omega)=R_0+{A}_0$ with $R_0$  
given by the continued fraction in (\ref{continued-fraction1}), then the
 zeros of $F(\omega)$ correspond to the characteristic values of $\omega$ for which
the condition (\ref{continued-fraction2}) is satisfied. In other words, $F(\omega)=0$ is
the eigenvalue equation for the radial eigenvalue problem.
Only for the denumerable infinite values of $\omega$ which are the roots of $F(\omega)=0$, do
we get solutions (\ref{power-series-soln1}) of the radial equation which are convergent at both $\tilde{r}=0$ and $\tilde{r}=r_H$.
The transcendental equation $F(\omega)=0$ may be solved for $\omega$ by standard root-search algorithms
(see e.g. \cite{Leaver86,Liu92} and references therein).

\section{Singular polynomial solutions to the radial equation}\label{polynomial}
In this section we study the existence of closed-form solutions of the radial equation which 
are polynomials in $r$ and thus automatically converge at $r=0, r_H$. 
Such solutions, if they exist, correspond to the values of $\omega$ which 
make the power series  (\ref{power-series-soln1}) truncate to become a polynomial of finite degree. 
%From the three-term recurrence relation, it
%can be seen that if for some positive integer $M$, $K_{M+1}=0$ and $\alpha=-(M-1)$, i.e.
Thus we seek solutions to (\ref{Heun-equation1}) which are of the form
\beq
X_M(\tilde{r})=\prod_{\ell=1}^M(\tilde{r}-\tilde{r}_\ell),\label{polynomial-ansatz}
\eeq
where $M=0,1,2,\cdots$ is the degree of the polynomial $X_M(\tilde{r})$,
$r_\ell$ are the roots of the polynomial and
 $X_M(\tilde{r})\equiv 1$ if $M=0$.
Closed-form (i.e. the so-called Liouvillian) solutions of the 
radial equation were discussed in \cite{Bini11} by means of the Kovacic algorithm \cite{Kovacic86}. 

We will follow the procedure proposed in \cite{Zhang12}. There exact
polynomial solutions of a general 2nd order linear differential equation were classified. 
Applying the results (e.g. Corollary 5.1 in the Appendix) of \cite{Zhang12},
we have that (\ref{polynomial-ansatz}) is a solution of the radial equation (\ref{Heun-equation1})
if the frequency $\omega$ and other system parameters satisfy the constraints
\beqa
\alpha\beta&=&-M(M-1)-2M(\alpha_0+\alpha^-_H+\alpha^+_H),\label{constraint1}\\
-q&=&-\left[2(M-1)+2(\alpha_0+\alpha^-_H+\alpha^+_H)\right]\sum_{\ell=1}^M(r_H-\tilde{r}_\ell)
      +2M(\alpha^+_H-\alpha^-_H)\,r_H,\label{constraint2}
 \eeqa
and the roots $\tilde{r}_\ell$ are determined by the set of $M$ algebraic equations
\beq
\sum_{\ell'\neq \ell}^M\frac{2}{\tilde{r}_\ell- \tilde{r}_{\ell'}}+\frac{2\alpha^-_H}{\tilde{r}_\ell}
   +\frac{2\alpha_0}{\tilde{r}_\ell-r_H}+\frac{2\alpha^+_H}{\tilde{r}_\ell-2r_H}=0,~~~~~~
   \ell=1,2,\cdots,M.\label{BEs}
\eeq
It is not hard to verify that (\ref{constraint1}) and  (\ref{constraint2}) can be simplified to
\beq
\alpha_0+\alpha^-_H+\alpha^+_H=-(M-1), ~~~~~~~~~2(M+\alpha_0)(\alpha^+_H-\alpha^-_H) =d_0 r_H,
    \label{constraint-combined}
\eeq
respectively and these two equations are equivalent for all $\alpha_0, \alpha^\pm_H$ values given 
in (\ref{alpha0-+}). 

It turns out that all closed-form solutions with the parameters satisfying the constraints
 (\ref{constraint-combined}) (equivalently (\ref{constraint1}) and (\ref{constraint2})) are singular at the poles of the radial equation. 
There are two classes (Class I and Class II) of singular polynomial solutions, 
classified according to whether the wave frequency $\omega$ has discrete or continuous spectra. An alternative approach of showing the existence of closed-form solutions is provided in the Appendix.
 
\subsection{Class I solutions - discrete wave frequency}
This class contains three families of solutions, which we describe as follows.

\vskip.1in
\noindent {\bf Class Ia.}  This  corresponds to the choice 
$\alpha_0=\frac{1}{2}\left(1+\sqrt{4\Lambda+1}\right),~~ 
\alpha^-_H=\frac{1}{2}(1+s-i\omega r_H),~~ \alpha^+_H=\frac{1}{2}(1-s-i\omega r_H)$. For this choice
the constraints (\ref{constraint-combined}) reduce to 
\beq
2M+1+\sqrt{4\Lambda+1}-2i\omega r_H=0.\label{constraint1a}
\eeq
This gives the discrete values of $\omega$
\beqa
\omega&=&-\frac{i}{2r_H}\left(2M+1+\sqrt{4\Lambda+1}\right)\n
&=&-\frac{i}{r_H}\left[N+M+1+\frac{1}{2}(|s-m|+|s+m|)\right],~~~~~~N, M=0,1,2,\cdots.
\eeqa
Here we have used (\ref{lambda}) for the separation constant $\Lambda$.
The corresponding solution $R(\tilde{r})$ is given by
\beqa
R(\tilde{r})&=&r^{s+1}_H\,(r_H-\tilde{r})^{\alpha_0-s-1}\,\tilde{r}^{\alpha^-_H-(s+1)/2}\,
   (2r_H-\tilde{r})^{\alpha^+_H-(s+1)/2}\,X_M(\tilde{r})\n
&=&r^{s+1}_H\,(r_H-\tilde{r})^{N-s+\frac{1}{2}(|s-m|+|s+m|)}\,\tilde{r}^{-\frac{i}{2}\omega r_H}\,
   (2r_H-\tilde{r})^{-s-\frac{i}{2}\omega r_H}\,\prod_{\ell=1}^M(\tilde{r}-\tilde{r}_\ell),
\eeqa
where $N, M=0,1,2,\cdots$ and $\tilde{r}_\ell$ are the solutions of the algebraic equations (\ref{BEs}). 
%This solution is divergent at $\tilde{r}=0$ (i.e. $r=r_H$). 

\vskip.1in
\noindent {\bf Class Ib.} This corresponds to the choice
 $\alpha_0=\frac{1}{2}\left(1-\sqrt{4\Lambda+1}\right),~~ 
\alpha^-_H=\frac{1}{2}(1+s-i\omega r_H),~~ \alpha^+_H=\frac{1}{2}(1-s-i\omega r_H)$. 
In this case, (\ref{constraint-combined}) reduce to
\beq
2M+1-\sqrt{4\Lambda+1}-2i\omega r_H=0,\label{constraint1c}
\eeq
which yields the characteristic values of $\omega$
\beq
\omega=-\frac{i}{r_H}\left[M-N-\frac{1}{2}(|s-m|+|s+m|)\right].
\eeq
The corresponding solution $R(\tilde{r})$ is
\beq
R(\tilde{r})=r^{s+1}_H\,(r_H-\tilde{r})^{-N-s-1-\frac{1}{2}(|s-m|+|s+m|)}\,
   \tilde{r}^{-\frac{i}{2}\omega r_H}\,
   (2r_H-\tilde{r})^{-s-\frac{i}{2}\omega r_H}\,\prod_{\ell=1}^M(\tilde{r}-\tilde{r}_\ell)
\eeq
with the roots $\tilde{r}_\ell$ determined by the algebraic equations (\ref{BEs}).
Here $N=0,1,\cdots$ and $M$ is an integer larger than or equal to $N+\frac{1}{2}(|s-m|+|s+m|)$, 
\beq
M=N+\frac{1}{2}(|s-m|+|s+m|),~~ N+\frac{1}{2}(|s-m|+|s+m|)+1, ~~\cdots
\eeq 
for stable or stationary solution (i.e. ${\rm Im}(\omega) \leq 0$). 
%This solution diverges at $\tilde{r}=r_H$ (i.e. $r=0$).

\vskip.1in
\noindent {\bf Class Ic.} This  corresponds to the choice
 $\alpha_0=\frac{1}{2}\left(1-\sqrt{4\Lambda+1}\right),~~ 
\alpha^-_H=\frac{1}{2}(1-s+i\omega r_H),~~ \alpha^+_H=\frac{1}{2}(1+s+i\omega r_H)$. 
In this case, (\ref{constraint-combined}) reduce to
\beq
2M+1-\sqrt{4\Lambda+1}+2i\omega r_H=0,\label{constraint1d}
\eeq
which yields the discrete values of $\omega$
\beq
\omega=-\frac{i}{r_H}\left[N+\frac{1}{2}(|s-m|+|s+m|-M)\right].
\eeq
The corresponding solution $R(\tilde{r})$ is given by
\beq
R(\tilde{r})=r^{s+1}_H\,(r_H-\tilde{r})^{-N-s-1-\frac{1}{2}(|s-m|+|s+m|)}\,
   \tilde{r}^{-s+\frac{i}{2}\omega r_H}\,
   (2r_H-\tilde{r})^{\frac{i}{2}\omega r_H}\,\prod_{\ell=1}^M(\tilde{r}-\tilde{r}_\ell),
\eeq
where the roots $\tilde{r}_\ell$ are computed from (\ref{BEs}). 
Here $N=0,1,\cdots$ and $M$ is an integer smaller than or equal to $N+\frac{1}{2}(|s-m|+|s+m|)$, 
\beq
M=0,1,\cdots, N+\frac{1}{2}(|s-m|+|s+m|)
\eeq
for stable or stationary solution. 
%This solution diverges at $\tilde{r}=r_H$ (i.e. $r=0$).

\subsection{Class II solutions - continuous wave frequency}
This class has two families of solutions, each has a continuous spectrum for the wave frequency $\omega$.
\vskip.1in
\noindent {\bf Class IIa.} This  corresponds to the choice
 $\alpha_0=\frac{1}{2}\left(1-\sqrt{4\Lambda+1}\right),~~ 
\alpha^-_H=\frac{1}{2}(1+s-i\omega r_H),~~ \alpha^+_H=\frac{1}{2}(1+s+i\omega r_H)$. 
In this case, (\ref{constraint-combined}) reduce to
\beq
2M+1-\sqrt{4\Lambda+1}+2s=0.\label{constraint1b}
\eeq
There is no constraint for $\omega$, i.e. $\omega$ takes continuous values. 
The corresponding solution $R(\tilde{r})$ reads
\beq
R(\tilde{r})=r^{s+1}_H\,(r_H-\tilde{r})^{-N-s-1-\frac{1}{2}(|s-m|+|s+m|)}\,
   \tilde{r}^{-\frac{i}{2}\omega r_H}\,
   (2r_H-\tilde{r})^{\frac{i}{2}\omega r_H}\,\prod_{\ell=1}^M(\tilde{r}-\tilde{r}_\ell),
\eeq
where $N=0,1,2,\cdots$ and $M=N-s+\frac{1}{2}(|s-m|+|s+m|)$; the roots $\tilde{r}_\ell$ are given
by the solutions of (\ref{BEs}). 
%This solution diverges at $\tilde{r}=r_H$ (i.e. $r=0$). 

\vskip.1in
\noindent {\bf Class IIb.} This  corresponds to the choice
 $\alpha_0=\frac{1}{2}\left(1-\sqrt{4\Lambda+1}\right),~~ 
\alpha^-_H=\frac{1}{2}(1-s+i\omega r_H),~~ \alpha^+_H=\frac{1}{2}(1-s-i\omega r_H)$. 
In this case, (\ref{constraint-combined}) reduce to
\beq
2M+1-\sqrt{4\Lambda+1}-2s=0.\label{constraint1e}
\eeq
The frequency $\omega$ is unconstrained and belongs to a continuous spectrum.
The corresponding solution $R(\tilde{r})$ reads
\beq
R(\tilde{r})=r^{s+1}_H\,(r_H-\tilde{r})^{-N-s-1-\frac{1}{2}(|s-m|+|s+m|)}\,
   \tilde{r}^{-s+\frac{i}{2}\omega r_H}\,
   (2r_H-\tilde{r})^{-s-\frac{i}{2}\omega r_H}\,\prod_{\ell=1}^M(\tilde{r}-\tilde{r}_\ell).
\eeq
Here $N=0,1,2,\cdots$ and $M=N+s+\frac{1}{2}(|s-m|+|s+m|)$; the roots $\tilde{r}_\ell$ are determined by
(\ref{BEs}). 
%This solution diverges at $\tilde{r}=r_H$ (i.e. $r=0$). 

\section{Summary and discussion}\label{summary}
We have provided a comprehensive study of the Teukolsky master equation for massless perturbations of 
any spin in de Sitter spacetime, and derived the analytic solutions for both the angular and radial 
parts of the equation. Furthermore, it is shown that the radial Teukolsky equation has no regular, 
closed-form solutions. We have presented two classes of singular polynomial solutions. The first class 
contains three families of solutions with discrete complex
frequency $\omega$, while the second class has two families of solutions with a continuous $\omega$.

\section*{Appendix}
In this Appendix, we provide an alternative approach of showing the existence of closed-form {\em singular} solutions for the radial equation with the system parameters satisfying the constraints (\ref{constraint-combined}). 

{}Imposing the constraint equations (\ref{constraint-combined}), we obtain from (\ref{alpha-beta-q}) that $\alpha=-(M-1)$, $\alpha\beta=M(M-1)$ and $-q=2M(\alpha^+_H-\alpha^-_H)r_H$. Substituting these into (\ref{A&B}) gives
\beqa
{A}_n&=&-\frac{(n-M)[3n-M+1+2(\alpha^-_H-\alpha^+_H)]}
   {2(n+1)(n+2\alpha^-_H)\,r_H}\n
%&=&-\frac{n(3n+2s+1)+L(4n+L+2s+1)}{2(n+1)(n+s+1-i\omega r_H)r_H},\n
{B}_n&=&\frac{(n-M)(n-M-1)}{2(n+1)(n+2\alpha^-_H)\,r^2_H}.\label{A&B-constrained}
%    =\frac{(n+L)(n+L-1)}{2(n+1)(n+s+1-i\omega r_H)r^2_H}.
\eeqa 
Thus when $n=M$, both $A_M$ and $B_M$ vanish. It follows from the three-term recurrence relation (\ref{three-term-relation1}) that $K_{M+1}=0$ and the power series (\ref{power-series-soln1}) truncates to give closed-form polynomial solutions in section \ref{polynomial}.  

Note that these polynomial solutions exist only when the system parameters (the frequency $\omega$ and other system parameters) satisfy the constraints (\ref{constraint-combined}). These constraints give the two classes of {\em singular} solutions presented in section \ref{polynomial}. As seen in section 5.2, for the class II solutions, although (\ref{constraint-combined}) do not provide any constraint on the parameter $\omega$, they do impose constraint between the value $M$ (i.e. the degree of the solution polynomial) and the other system parameters $\Lambda$ and $s$, see (\ref{constraint1b}) and (\ref{constraint1e}).

We also remark that for the parameters in (\ref{alpha0-+nonsingular}) corresponding to {\em nonsingular} solutions, the constraint equations (\ref{constraint-combined}) give rise to $2M+1+\sqrt{4\Lambda+1}+2s=0$, i.e. $M=-N-s-1-\frac{1}{2}(|s-m|+|s+m|)$. That is, $M$ is negative. Thus there are no {\em nonsingular} polynomial solutions for the radial equation and solutions for this case are given by power series in section \ref{radial}.

\section*{Acknowledgement}
This work was partially supported by the Australian Research Council through Discovery-Projects grant DP190101529.

\section*{Data Availability Statement}
The data that support the findings of this study are available within the article.

\end{document}